# Reverse Aperiodic Resonance in Low- to High-Dimensional Bistable Systems: A Complement to Stochastic Resonance Studies in Logic Circuits

Mengen Shen[1], Jianhua Yang[1*], Miguel A. F. Sanjuán[2], Huatao Chen[3], and Zhongqiu Wang[4]

[1] Jiangsu Key Laboratory of Mine Mechanical and Electrical Equipment, School of Mechanical and Electrical Engineering, China University of Mining and Technology, Xuzhou 221116, Jiangsu, People's Republic of China

[2] Nonlinear Dynamics, Chaos and Complex Systems Group, Departamento de Física, Universidad Rey Juan Carlos, Tulipán s/n, 28933 Móstoles, Madrid, Spain

[3] Division of Dynamics and Control, School of Mathematics and Statistics, Shandong University of Technology, Zibo 255000, People's Republic of China

[4] School of Computer Science and Technology, China University of Mining and Technology, Xuzhou 221116, Jiangsu, People's Republic of China

**Abstract**

As circuits continue to miniaturize, noise has become a significant obstacle to performance optimization. Stochastic resonance in logic circuits offers an innovative approach to harness noise constructively; however, current implementations are limited to basic logical functions such as OR, AND, NOR, and NAND, restricting broader applications. This paper introduces a three-dimensional (3D) coupling model to investigate the counterintuitive phenomena that arise in nonlinear systems under noise. Compared to the one-dimensional Langevin equation and the two-dimensional Duffing equation, the 3D coupling model features more adjustable parameters and coupling interactions, enhancing the system's dynamic behavior. The study demonstrates that increasing noise intensity triggers reverse aperiodic resonance, leading to signal phase reversal and amplitude amplification. This phenomenon is attributed to the motion of Brownian particles in a bistable potential well. Additionally, reverse aperiodic resonance addresses the lack of logical negation in traditional stochastic resonance systems by introducing noise-driven phase reversal, providing a novel alternative to conventional inverters.

* Corresponding author. jianhuayang@cumt.edu.cn

# 1. Introduction

As the trend toward circuit miniaturization accelerates, noise has emerged as a significant challenge to circuit performance [1]. In logic circuits, where precision and stability are critical, noise is typically regarded as detrimental to signal processing. However, recent advances in the field of stochastic resonance (SR) have revealed the potential for noise to play a beneficial role in enhancing signal transmission and processing [2, 3]. SR has been explored in various physical systems as a means to amplify weak signals through the constructive use of noise [4-6]. In logic circuits, SR has introduced a new dimension to noise utilization, enabling the execution of logic operations such as OR, AND, NOR and NAND with noise assistance [7, 8]. However, the current scope of SR in logic circuits remains constrained by the inability to perform more advanced operations, which limits its application to more complex computational tasks [9, 10].

One key limitation is that current SR-based technologies cannot implement the logical NOT operation [11]. As an essential component of digital circuits, logical NOT plays a crucial role in enabling complex logic functions [12, 13]. Without its inclusion, the practical utility of SR in logic circuits is greatly restricted. Comprehensive logic design requires the generation of all logical operations, including XOR and XNOR, from fundamental gates like OR, AND, and NOT [14-16]. Additionally, the counterintuitive dynamics of SR, especially in nonlinear systems, remains insufficiently explored, which hampers the full potential of noise-driven phenomena in advancing innovative circuit designs [17, 18].

To address these limitations, this paper introduces a three-dimensional (3D) coupling model that broadens the scope of SR in nonlinear systems by incorporating additional degrees of freedom and coupling effects. Unlike the one-dimensional Langevin equation and the two-dimensional Duffing equation, the 3D coupling model introduces more adjustable parameters, which enriches the dynamic behavior and expands the range of noise-induced phenomena [19-22]. This model provides a deeper investigation into reverse aperiodic resonance, a lesser-known phenomenon where increasing noise strength causes phase reversal and amplitude amplification in the system's response [23, 24]. Reverse aperiodic resonance differs from traditional resonance behaviors, adding a layer of complexity with significant implications for signal processing and noise management in various environments.

The mechanism behind this counterintuitive phenomenon can be explained by the movement of Brownian particles within a bistable potential well [25, 26]. The interaction between particle motion and the bistable potential leads to phase shifts and amplitude changes, particularly in systems modulated by external signals [27, 28]. This new understanding not only enhances the theoretical foundations of SR but also paves the way for engineering applications in environments where noise is inevitable. A key contribution of this study is the demonstration that reverse aperiodic resonance can be harnessed to achieve logical negation in SR systems. By exploiting noise-driven phase reversal, the proposed model overcomes the absence of logical NOT in traditional SR logic systems, presenting a novel approach to noise-resistant circuit design. In summary, this work advances the theoretical and practical understanding of SR in logic circuits by addressing the missing logical NOT function. The introduction of the 3D coupling model enriches the dynamic behaviors that can be explored in SR systems and opens new avenues for noise-assisted computing, signal processing, and control systems.

The remainder of this paper is organized as follows: In Sect. 2, the 3D coupling model is introduced, including an analysis of equilibrium points and their stability. Additionally, a comparison of bistable nonlinear systems across different dimensions (from low to high) is provided. A comparative analysis of three typical signals and four types of noise before and after processing by the three bistable models is offered in Sect. 3. This section further explores the mechanism behind reverse aperiodic resonance by examining the transitions of Brownian particles between bistable potential wells. Section 4 discusses potential real-world applications of reverse aperiodic resonance, particularly its role as a substitute for inverters in logic circuits, thereby addressing the gap in logical SR. A thorough discussion on the results is presented in Sect. 5. Finally, Sect. 6 concludes the study by summarizing the main findings.

## 2. Three nonlinear models

### 2.1 Langevin equation (1D)

The bistable Langevin equation is one of the fundamental models for studying SR, commonly used to describe the dynamics of systems with bistable characteristics under the influence of noise

$$\frac{dx(t)}{dt} = ax(t) - bx^3(t) + s(t) + \xi(t), \qquad (1)$$

where $x$ is the system state variable and $ax(t)$ is a linear restoring force term, driving the system back to the origin, while $-bx^3(t)$ is a nonlinear term, generating bistable behavior, allowing the system to switch between two stable points. The function $s(t)$ is an external drive, representing the influence of the external forces, while $\xi(t)$ is the noise term, representing random disturbances. This is a single variable system that describes the dynamic behavior of only one variable $x(t)$.

## 2.2 Duffing equation (2D)

The Duffing equation is one of the classic nonlinear vibration equations, often used to describe nonlinear systems with bistable characteristics. Its mathematical standard form is given by

$$\frac{dx^2(t)}{dt^2} + \mu \frac{dx(t)}{dt} + Ax(t) + Bx^3(t) = s(t) + \xi(t), \qquad (2)$$

where $\mu dx/dt$ is the linear damping term, describing the energy loss in the system. The function $Ax(t)$ is the linear restoring force, driving the system back to equilibrium, while $Bx^3(t)$ is the nonlinear restoring force, generating bistability. The function $s(t)$ is the external driving force, while $\xi(t)$ is the noise term, representing random disturbances.

By introducing $y = dx/dt$, the equation can be decomposed into two first-order differential equations

$$\begin{cases} \dfrac{dx}{dt} = y \\ \dfrac{dy(t)}{dt} = Ax(t) - Bx^3(t) + \mu y(t) + s(t) + \xi(t) \end{cases}. \qquad (3)$$

SR can occur under the combined influence of the external driving signal $s(t)$ and noise $\xi(t)$. Since this equation is two-dimensional, it can describe more complex phase space behaviors such as transitions between bistable states, periodic orbits, and chaotic phenomena.

## 2.3 3D coupling model

Most models used in research share the common feature of low dimensionality. Although the mathematical expressions of low-dimensional models are concise, they offer fewer adjustable parameters, limiting their flexibility. This paper proposes a 3D model that generalizes the equation such that the low-dimensional equation becomes a special case of the degenerate high-dimensional form

$$\begin{cases} \dfrac{dx}{dt} = \alpha x - \kappa xz + s(t) \\ \dfrac{dy}{dt} = \sigma(x - y) \\ \dfrac{dz}{dt} = -\gamma z + \rho xy \end{cases}, \quad (4)$$

where the second equation $dy/dt=\sigma(x-y)$ is similar to the equation of the classical Lorenz system, describing the interaction between $x$ and $y$. The first equation $dx/dt=\alpha x-\kappa xz+s(t)$ involves the coupling with the variable $z$ and an external driving term $s(t)$, and the third equation $dz/dt=-\gamma z+\rho xy$ describes the coupling between variables $x$ and $y$.

The dimensions and parameters of the new model are presented in Eq. (4). The model is highly flexible and can be tailored to various signal processing needs, leveraging its multiple adjustable parameters. This system of 3D equations can be seen as a coupling system with higher-dimensional complexity. The coupling terms $\kappa xz$ and $\rho xy$ in each equation describe the interactions between different state variables.

Counterintuitive phenomena are intricately associated with changes in the dynamic behavior of nonlinear systems under the influence of external input signals. Accordingly, it is crucial to conduct a thorough dynamic analysis of the 3D equations outlined in Eq. (4). Next, the dynamic processes will be investigated by analyzing the stability of equilibrium points and the characteristics of attractor trajectories.

**2.3.1 Equilibrium points**

To find the equilibrium point of the 3D coupling model, we set $dy/dt = dz/dt = 0$ and solve for the equilibrium points of the system

$$\begin{cases} 0 = \sigma(x - y) \\ 0 = -\gamma z + \rho xy \end{cases}, \quad (5)$$

so that we obtain,

$$\begin{cases} y = x \\ z = \dfrac{\rho}{\gamma} x^2 \end{cases}. \quad (6)$$

There is always $\dot{x} = 0$ when the system is at the equilibrium point, so that,

$$\frac{dx}{dt} = \alpha x - \frac{\kappa \rho}{\gamma} x^3 + s(t) = 0. \quad (7)$$

Let $f_{eq} = -\alpha x + \frac{\kappa\rho}{\gamma}x^3$, $\alpha < 0$, $\frac{\kappa\rho}{\gamma} < 0$. The intersections of $f_{eq}$ and $s(t)$ are equilibrium points. As shown in Fig. 1, where $M$, $O$, $N$ are intersections of $f_{eq}(x)$ and $s(t)=0$, and $P$, $Q$ are extreme points of $f_{eq}(x)$. The points $I_1$, $I_2$, $I_3$ are intersections of $s(t)=m$ and $f_{eq}(x)$, i.e., the equilibrium points of system. According to derivation, the coordinate expression of each key point are $M\left(-\sqrt{\frac{\alpha\gamma}{\kappa\rho}}, 0\right)$, $O(0,0)$, $N\left(\sqrt{\frac{\alpha\gamma}{\kappa\rho}}, 0\right)$, $P\left(-\sqrt{\frac{\alpha\gamma}{3\kappa\rho}}, \sqrt{\frac{4\alpha^3\gamma}{27\kappa\rho}}\right)$, and $Q\left(\sqrt{\frac{\alpha\gamma}{3\kappa\rho}}, -\sqrt{\frac{4\alpha^3\gamma}{27\kappa\rho}}\right)$.

When the external input $s(t) \in \left(-\sqrt{\frac{4\alpha^3\gamma}{27\kappa\rho}}, \sqrt{\frac{4\alpha^3\gamma}{27\kappa\rho}}\right)$, there are always three intersection points between the external input $s(t)$ and $f_{eq}(x)$. Next, we will analyze the stability of each equilibrium point.

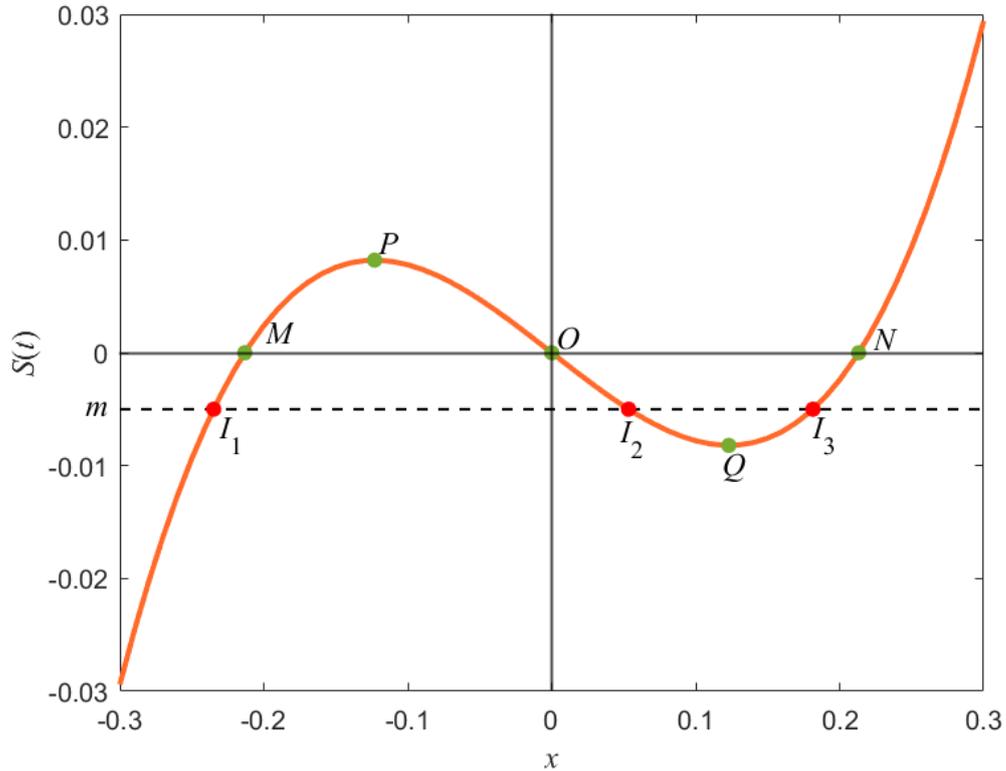

**Fig. 1.** Equilibrium points of 3D coupling model. The distribution of equilibrium points can be analyzed using bifurcation curves from the bistable potential function. Without external input signals, the equilibrium points $M$, $O$, and $N$ are the intersections of the horizontal axis and the curve. With external signals, new equilibrium points emerge at the intersections; for instance, when $s(t)=m$, the points are $I_1$, $I_2$, and $I_3$

### 2.3.2 Stability analysis of equilibrium points

Using the Jacobian matrix, we can calculate the eigenvalues at the equilibrium point to analyze the stability of the system

$$J = \begin{bmatrix} \alpha & -\kappa z & -\kappa x \\ \sigma & -\sigma & 0 \\ \rho y & \rho x & -\gamma \end{bmatrix}. \tag{8}$$

Let $|J-\lambda I|=0$, where $I$ is the identity matrix, $\lambda=p+qi$. Combining Eq. (5) and Eq. (8), we can see that the real part $p$ and imaginary part $q$ of the three eigenvalue $\lambda$ varies with $x$ shown in Fig. 2.

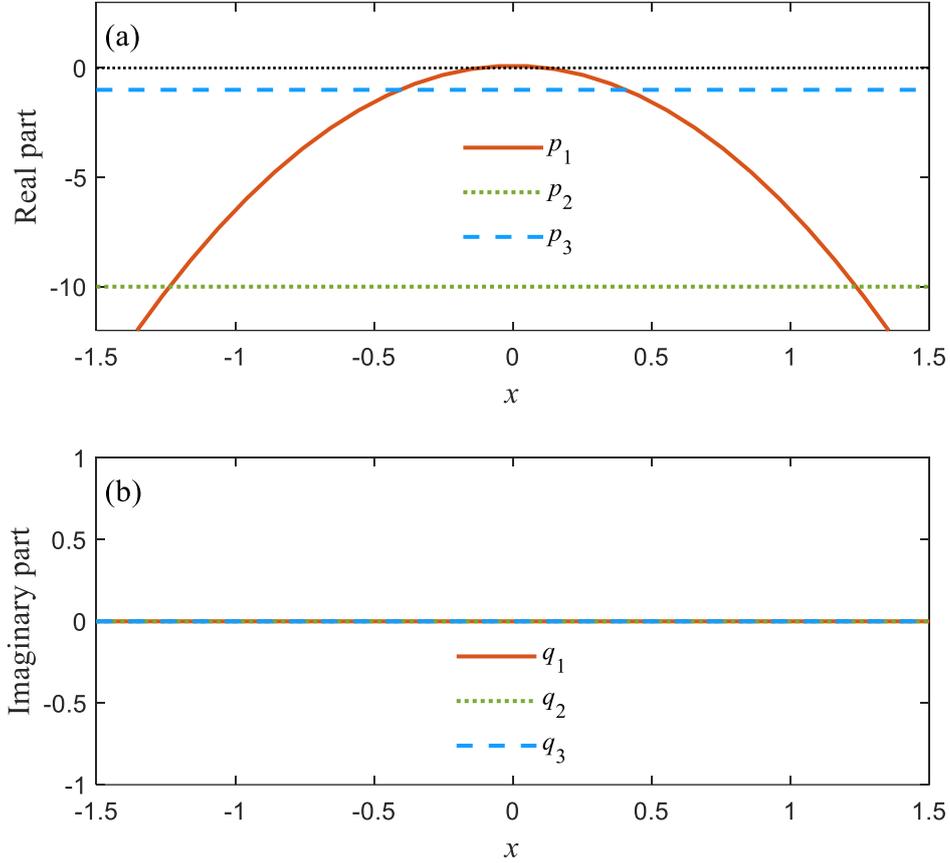

**Fig. 2.** Eigenvalues of the 3D coupling model. The stability of the nonlinear system can be intuitively assessed by the signs of the real and imaginary parts of the eigenvalues. Positive real parts indicate instability, while negative real parts suggest stability. The behavior of the imaginary parts offers insights into oscillatory dynamics. (a) Real parts, (b) Imaginary parts. Parameters: $\sigma=10$, $\rho=-0.1$, $\alpha=-2.2$, $\gamma=1$, $\kappa=1$

When $x \in [-\infty, -\sqrt{\frac{\alpha\gamma}{\kappa\rho}}] \cup [\sqrt{\frac{\alpha\gamma}{\kappa\rho}}, +\infty]$, the real parts of the eigenvalues $p_1$, $p_2$, $p_3$ are all negative numbers. If there are equilibrium points within this range, it is stable, meaning that the dynamic behavior of the system will gradually approach these equilibriums. In addition, when $x \in [-\sqrt{\frac{\alpha\gamma}{\kappa\rho}}, \sqrt{\frac{\alpha\gamma}{\kappa\rho}}]$, there is always a positive $p_1$. It is known that when the real part of the eigenvalue is positive, the equilibrium point is unstable, which means that the system will diverge from the equilibrium point and move away from the equilibrium point [29]. That is, a small perturbation will cause the system to move away from the equilibrium point.

The imaginary part of the eigenvalue determines whether the system exhibits oscillatory behavior. It can be seen from Fig. 2(b) that the imaginary parts of the three eigenvalues are all zero, that is, the system has no oscillation behavior. In summary, the two outer sides of the three equilibrium points are stable equilibrium points, and the middle is an unstable equilibrium point, which constitutes a bistable structure.

**2.4 Comparison of different bistable models**

To illustrate the differences among bistable models from low to high dimensional, Table 1 provides a comparison between the traditional one- and two-dimensional models and the 3D coupling model introduced in this paper.

In Table 1, $s(t) = s(t) + \xi(t)$ represents the input signal of the nonlinear system, where $s(t)$ is the signal to be processed and $\xi(t)$ is the noise signal. The table shows that the equivalent curves of equilibrium points and thresholds across the three bistable models, from low to high dimensional, are essentially the same. However, compared to the 1D Langevin and 2D Duffing models, the 3D coupling model offers the following major advantages: *A. More adjustable parameters.* The Langevin equation has only two adjustable parameters, *a* and *b*, while the Duffing equation also has only two adjustable parameters, *A* and *B*, which limits the flexibility of the system. In contrast, the 3D coupling model introduces four parameters in the expression of the equilibrium point curve, $\rho$, $\alpha$, $\gamma$ and $\kappa$, significantly enhancing the flexibility of parameter adjustment. In fact, all five parameters in the 3D coupling model can be adjusted, providing a more comprehensive and flexible system tuning approach. *B. More useful system outputs.* Based on the dynamic analysis of the nonlinear model, the output of the bistable model is generated only in a single *x*-direction. There is no essential difference between the Langevin model and the Duffing model when the same input signal $s(t)$ is applied to the system. However, in the 3D coupling model, when noise causes counterintuitive phenomena, the relationship between the state variables $x$, $y$, and $z$ in Eq. (4) always remains consistent. This means signals of different amplitudes can be generated across different dimensions, greatly facilitating signal processing. For instance, when the parameters are set, the output signal amplitudes in the $y$ and $z$ directions are $A_y = A_x$, $A_z = \frac{\rho}{\gamma} A_x^2$. Therefore, using the proposed 3D coupling model, three signals of different amplitudes can be generated, and these amplitudes can be flexibly adjusted by tuning the parameters.

**Table 1.** Comparison of low- to high-dimensional bistable nonlinear systems

| Bistable model | 1D Langevin equation | 2D Duffing equation | 3D coupling Model |
|---|---|---|---|
| Mathematical Expression | $\frac{dx}{dt} = ax - bx^3 + s(t)$ | $\frac{d^2x}{dt^2} + \mu\frac{dx}{dt} = Ax - Bx^3 + s(t)$ | $\begin{cases} \frac{dx}{dt} = \alpha x - \kappa xz + s(t) \\ \frac{dy}{dt} = \sigma(x - y) \\ \frac{dz}{dt} = -\gamma z + \rho xy \end{cases}$ |
| System Dynamics | 1D bistable system driven by a cubic nonlinearity, creating two stable states. | 2D nonlinear oscillator with damping and restoring forces, producing complex bistable behavior. | 3D coupling nonlinear system showing rich dynamics, including bistability. |
| Equilibrium points Equivalent curve | $f_{eq} = -ax + bx^3$ | $f_{eq} = -Ax + Bx^3$ | $f_{eq} = -\alpha x + \frac{\kappa\rho}{\gamma} x^3$ |
| Thresholds | $\sqrt{\frac{4a^3}{27b}}$ | $\sqrt{\frac{4A^3}{27B}}$ | $\sqrt{\frac{4\alpha^3\gamma}{27\kappa\rho}}$ |
| Adjustable Parameters | $a, b$ | $A, B, \mu$ | $\sigma, \rho, \alpha, \gamma, \kappa$ |

# 3. Reverse aperiodic resonance

## 3.1 Typical circuit signals

The mathematical representation of signals in different types of circuits varies. Digital signals are discrete binary signals that usually represent high and low levels of logic levels, that is, logic 1 and logic 0. Its mathematical expression commonly uses Boolean algebra to express the relationship between signals, or describes the changes of signals through time series. Depending on their characteristics, digital signals can be divided into the following types [30].

### 3.1.1 Square wave signal (clock signal)

In digital circuits, clock signals are typically square wave signals. The mathematical expression for a clock signal is,

$$x(t) = A \cdot \text{sgn}(\sin(2\pi f t)), \tag{9}$$

where $A$ is the amplitude of the signal, typically 1 or 0. The frequency of the signal is $f$, representing the switching speed, while $\text{sgn}(\sin(2\pi f t))$ is the sign function, indicating the transition between high and low logic levels within each cycle.

### 3.1.2 Pulse train signal

In digital circuits, a pulse train signal can be represented as a series of pulses. The mathematical expression is,

$$x(t) = \sum_{n=-\infty}^{\infty} A_n \cdot \delta(t - nT), \tag{10}$$

where $A_n$ is the amplitude of each pulse (either 1 or 0). $\delta(t-nT)$ is the Dirac delta function, indicating the position of each pulse, and $T$ is the period of the pulse train.

### 3.1.3 Step signal

A step signal represents a transition to a high logic level at a specific point in time and is often used in switching circuits. Its mathematical expression is

$$x(t) = A \cdot u(t - t_0), \tag{11}$$

where $A$ is the amplitude of the step signal, and $u(t-t_0)$ is the unit step function, where the signal is 0 before $t_0$ and transitions A after $t_0$. To provide a visual representation, we illustrate the waveforms of these different types of circuit signals in Fig. 3. The responses of different models to these types of circuit signals can be observed in varying conditions. It should be pointed out that to facilitate the processing of bistable nonlinear systems, the three typical circuit signals here are uniformly polarized.

It is worth explaining that typical circuit signals include periodic signals and aperiodic signals. Aperiodic signals refer to signals that are not repetitive in time, while periodic signals are signals that repeat within a specific time interval. Therefore, periodic signals are actually special types of aperiodic signals.

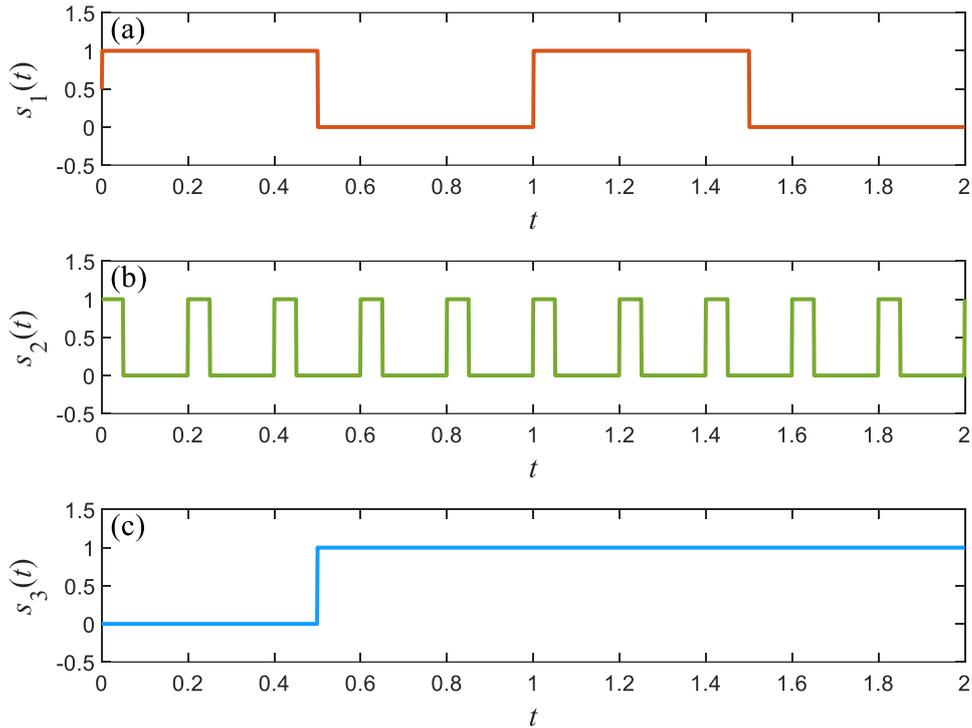

**Fig. 3.** Three typical waveforms of binary aperiodic signals in digital circuits. (a) Square wave signal (clock signal), (b) pulse sequence signal,(c) step signal

### 3.2 Typical circuit noise

Digital signals in circuits are subject to various types of noise, which can affect signal integrity and circuit performance. Below are some common types of noise found in digital circuits, along with brief explanations [31].

### 3.2.1 Thermal noise

Thermal noise caused by the random motion of electrons in a conductor due to heat, is an unavoidable type of noise in all circuits. It is often used to describe voltage or current noise in resistors, capacitors, or inductors. This noise is related to temperature, with its power increasing as temperature rises. While primarily affecting analog and low-power circuits, thermal noise can also influence high-speed digital signals

$$v_n(t) = \sqrt{4k_B T R \Delta f} \cdot N(t), \tag{12}$$

where $k_B$ is Boltzmann constant ($1.38 \times 10^{-23}$ J/K), $T$ is the temperature (Kelvin), $R$ is the resistance (Ohms), $\Delta f$ is the bandwidth (Hertz), and $N(t)$ is Gaussian white noise with zero mean and unit variance.

### 3.2.2 Shot noise

Shot noise is caused by the discrete motion of charge carriers across a PN junction or similar component. It is a type of random noise that is proportional to the current passing through the component

$$i_n(t) = \sqrt{2qI\Delta f} \cdot N(t), \tag{13}$$

where $q$ is the electron charge ($1.6 \times 10^{-19}$ C), $I$ is the current (Amperes), $\Delta f$ is the bandwidth (Hertz), and $N(t)$ is Gaussian white noise.

### 3.2.3 Flicker noise

Flicker noise, also known as $1/f$ noise, becomes more prominent at lower frequencies. It is common in many semiconductor devices and increases in intensity as frequency decreases

$$v_f(t) = \frac{K_f}{f\gamma} \cdot N(t), \tag{14}$$

where $K_f$ is a constant related to the device characteristics, $f$ is the frequency, $\gamma$ is the frequency decay factor, typically close to 1, and $N(t)$ is Gaussian white noise.

### 3.2.4 Electromagnetic interference

Electromagnetic interference or power supply noise is caused by external noise sources such as power fluctuations or radio waves. This noise can be periodic or random and is often simulated by adding oscillating or low-frequency signals

$$v_{emi}(t) = A\sin(2\pi f_{emi} t + \phi) + N(t), \tag{15}$$

where $A$ is the amplitude of the power supply noise, $f_{emi}$ is the frequency of the electromagnetic interference, $\phi$ is the initial phase, and $N(t)$ is Gaussian white noise representing the random component.

These noise types are common sources of interference in digital circuits, and they must be managed to ensure stable circuit performance.

## 3.3 Comparison of signals before and after processing by bistable models

For signals in circuits, the correlation coefficient can be employed as a metric to assess the signal before and after bistable model processing. According to the physical interpretation of the correlation coefficient, a larger value indicates a greater similarity between the two time series. When the correlation coefficient equals 1, the two time series are completely similar. Conversely, when the correlation coefficient equals -1, the inverted processed signal is completely similar to the signal before processing. In other words, the greater the absolute value of the correlation coefficient, the more similar the signals before and after processing are. A negative correlation coefficient signifies inversion. Its specific calculation expression is

$$C_{sx} = \frac{\sum_{i=1}^{k}[s(i)-\bar{s}][x(i)-\bar{x}]}{\sqrt{\sum_{i=1}^{k}[s(i)-\bar{s}]^2[x(i)-\bar{x}]^2}}, \tag{16}$$

where $\bar{s}$ and $\bar{x}$ represent the mean values of the signal before bistable model processing and the signal after processing, respectively.

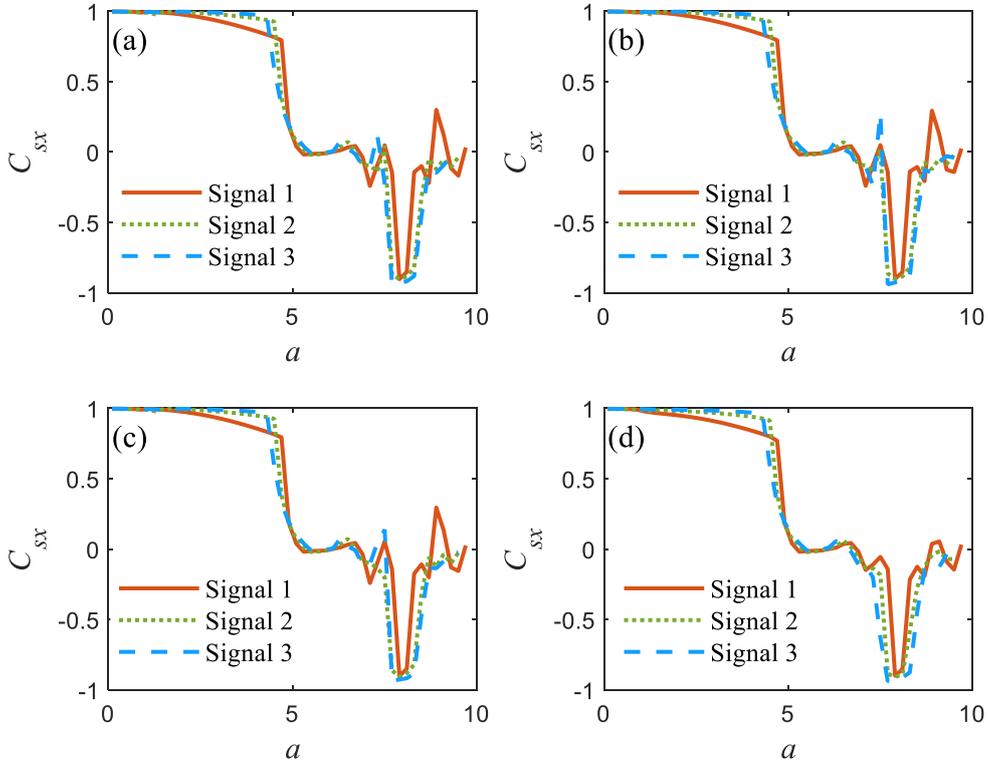

**Fig. 4.** Correlation between the response of Langevin equation (1D bistable nonlinear system) and three typical inputs under different types of noise as a function of the system parameter. (a) Thermal noise, (b) shot noise, (c) flicker noise, (d) electromagnetic interference

Figure 4 illustrates the correlation $C_{sx}$ between the response of the Langevin equation (a 1D bistable nonlinear system) and three typical input signals under different noise conditions. The figure demonstrates how the system exhibits an inverse correlation under the influence of thermal, shot, flicker, and electromagnetic noise. In this case, the processed signal is inversely related to the input signal, which means that the output signal of the system is inverted relative to the input. This phenomenon is indicated by the correlation coefficient falling below zero, indicating that the system response reaches its maximum negative correlation upon reversal. This behavior highlights novel effects of inverse correlation in the presence of noise.

The results reveal a consistent pattern in which systems experience strong inverse correlations under the influence of noise, suggesting that noise can play a unique and critical role in altering system responses.

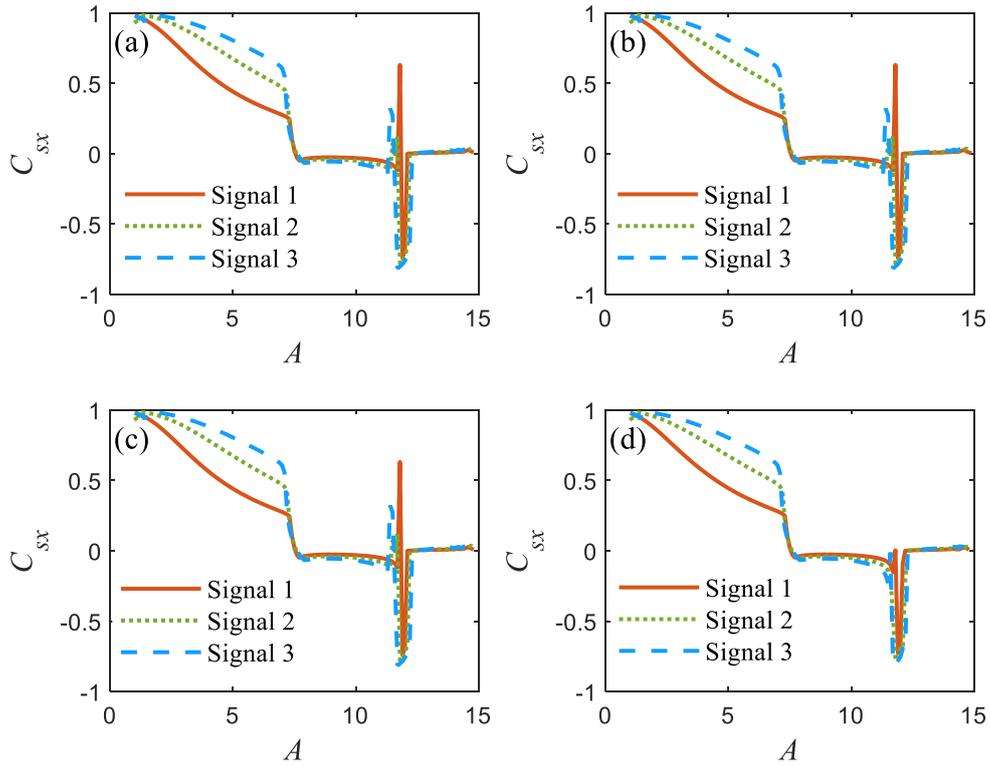

**Fig. 5.** Correlation between the response of the Duffing equation (2D bistable nonlinear system) and three typical inputs under different types of noise as a function of the system parameter. (a) Thermal noise, (b) shot noise, (c) flicker noise, (d) electromagnetic interference

The next step is to explore whether this inverse correlation phenomenon can also be observed in a two-dimensional bistable system, modeled by the Duffing equation. Figure 5 demonstrates the correlation $C_{sx}$ between the response of the Duffing equation (a 2D bistable nonlinear system) and three typical input signals under different noise conditions. As in the previous figure, the system exhibits a strong inverse correlation with the input signals across various noise types. The correlation coefficient reaches its most negative value, indicating that the processed signal is maximally correlated with the inverted input signal. This reinforces the observation of this distinctive inverse correlation behavior in bistable systems under noise influence.

The results confirm that, like the 1D system, the 2D bistable system also shows a pronounced inverse correlation under the influence of noise, demonstrating that this behavior persists as system complexity increases. Following this, the model is extended from the 2D bistable system to the 3D coupling bistable system, allowing us to examine whether this inverse correlation phenomenon remains observable in higher-dimensional systems.

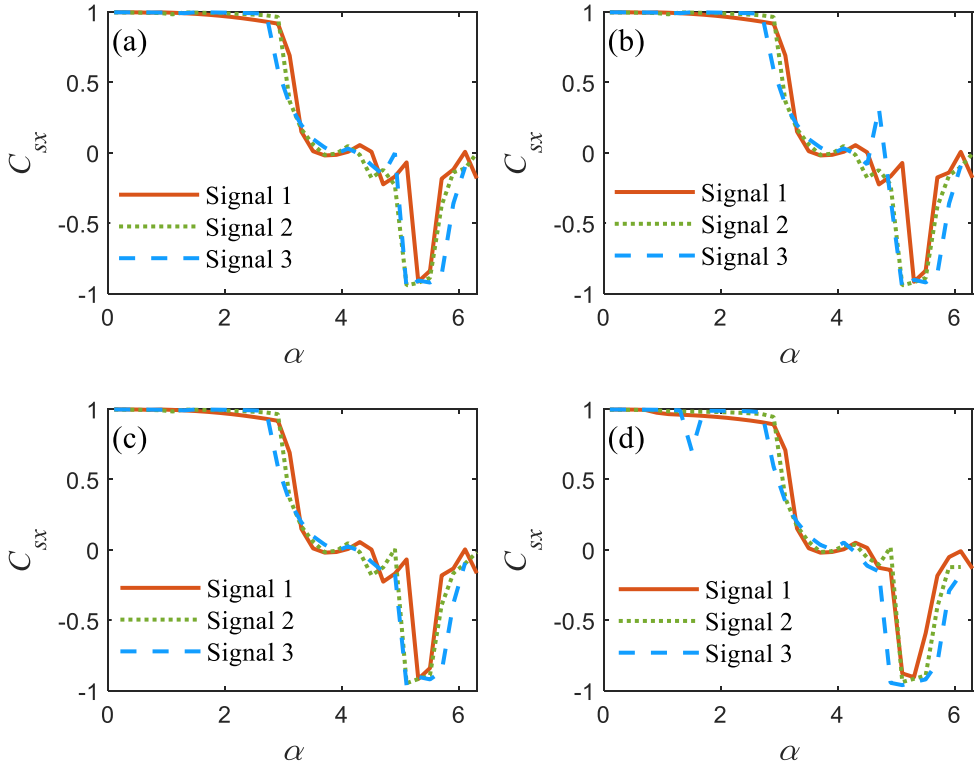

**Fig. 6.** Correlation between the response of the 3D coupling bistable nonlinear system and three typical inputs under different types of noise as a function of the system parameter. (a) Thermal noise, (b) shot noise, (c) flicker noise, (d) electromagnetic

interference

Figure 6 presents the correlation $C_{sx}$ between the response of the 3D coupling bistable nonlinear system and three typical input signals under different noise conditions. The pattern observed in the previous figures remains consistent: the system exhibits inverse correlation with the input signals under various noise influences. The correlation coefficient again reaches negative values, indicating that the processed signal is maximally correlated with the inverted input signal.

This result emphasizes the novel behavior of inverse correlation in bistable systems when noise is present, illustrating that this phenomenon persists even as the system complexity increases. The trend remains consistent, showcasing similar effects across 1D, 2D, and 3D systems.

The findings reinforce the idea that the inverse correlation effect is a robust feature of bistable systems affected by noise, regardless of the dimensionality or complexity of the system.

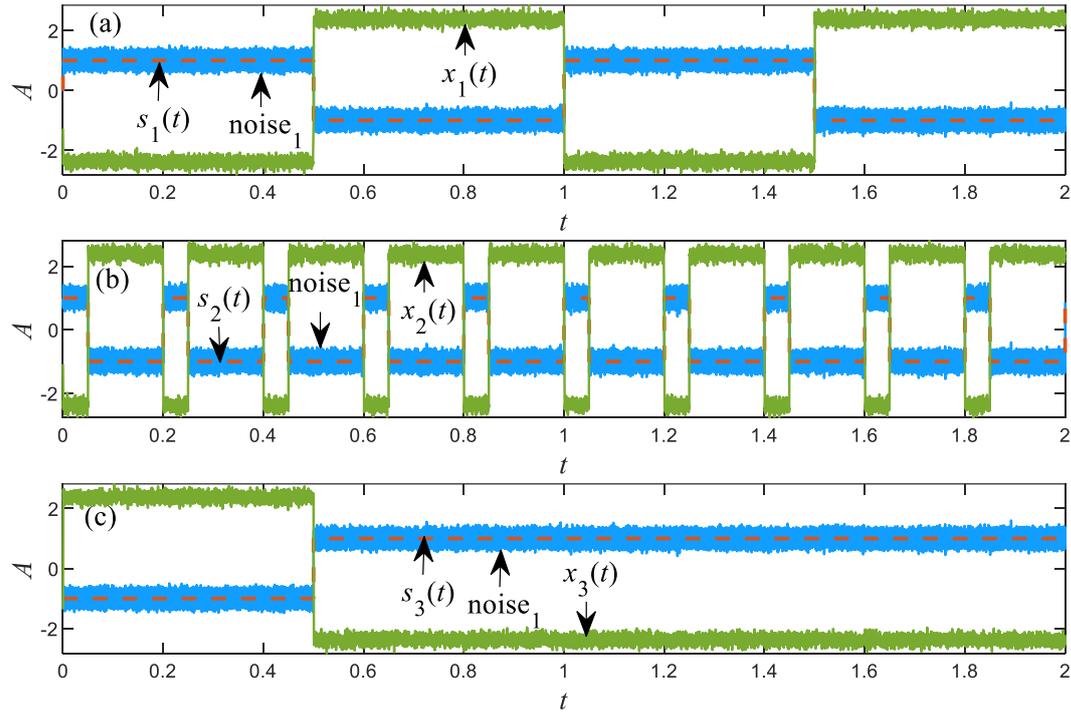

**Fig. 7.** Comparison of waveforms before and after processing three typical signals through the 3D coupling Model under the influence of the same noise. (a) Square wave signal (clock signal), (b) pulse sequence signal, (c) step signal

Next, observe the behavior when the correlation coefficient reaches its maximum negative value, and evaluate how the signal is processed and whether the processed signal aligns with the inverted input signal. To validate this, Fig. 7 shows a comparison of the three typical signals processed through the 3D coupling model under the influence of the same noise. As illustrated in the figure, when noise is introduced, the processed signal exhibits fluctuations and phase inversions, demonstrating consistent behavior across all signals. This novel phenomenon highlights the emergence of inverse correlation under noise influence.

Taking the analysis further, Fig. 8 compares the same input signal processed through three different bistable nonlinear systems under identical noise conditions. As shown in the figure, the processed signal in each system (1D Langevin, 2D Duffing, and 3D coupling model) experiences significant fluctuations and phase inversions like those observed in previous cases. The correlation behavior remains consistent across all systems, once again demonstrating the same novel inverse correlation effect in response to noise.

Further expanding this observation, Fig. 9 provides a comparison of waveforms before and after processing the same signal through the 3D coupling model, but this time under the influence of four different types of noise: thermal noise, shot noise, flicker noise, and electromagnetic interference. In each case, the system exhibits similar behavior, where the processed signal shows noticeable amplitude fluctuations and phase inversions, correlating with the noise-induced disruptions. This consistent inverse relationship across different noise types further substantiates the robustness of this novel inverse correlation phenomenon.

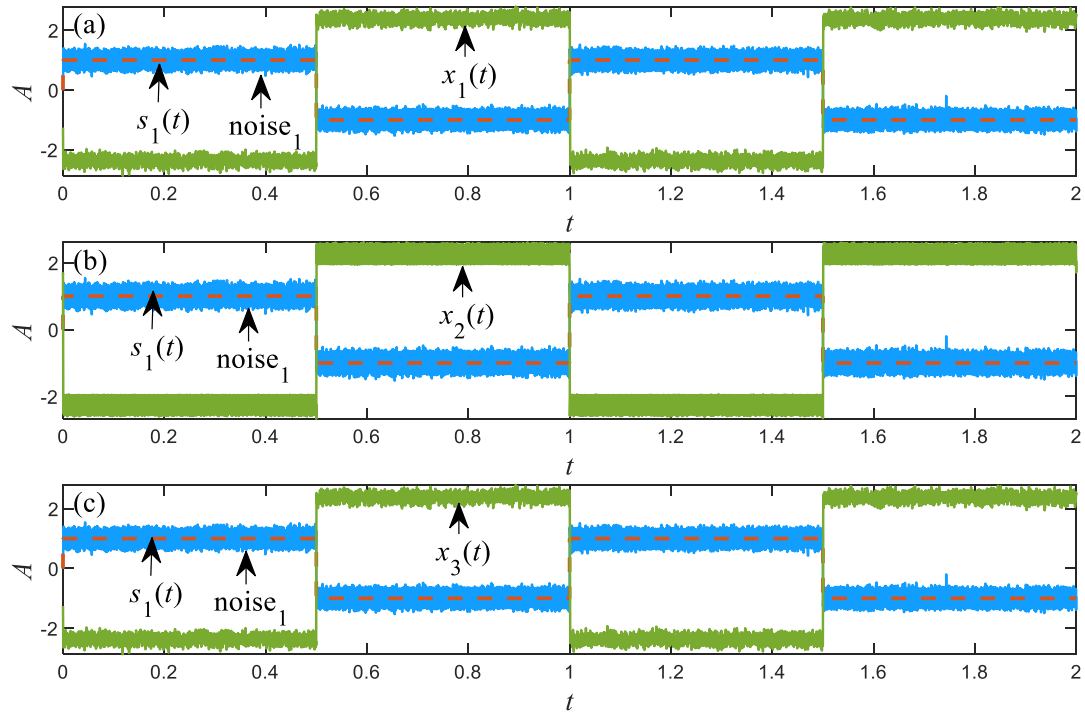

**Fig. 8.** Comparison of waveforms before and after processing the same signal through three bistable nonlinear systems under the influence of the same noise. (a) 1D Langevin equation, (b) 2D Duffing equation, (c) 3D coupling model

The results confirm that the inverse correlation behavior observed earlier is not limited to a specific noise type or system complexity but is a consistent characteristic of bistable nonlinear systems under varying noise influences. This behavior represents a significant insight into the dynamic response of bistable systems and their resilience to noise-induced perturbations.

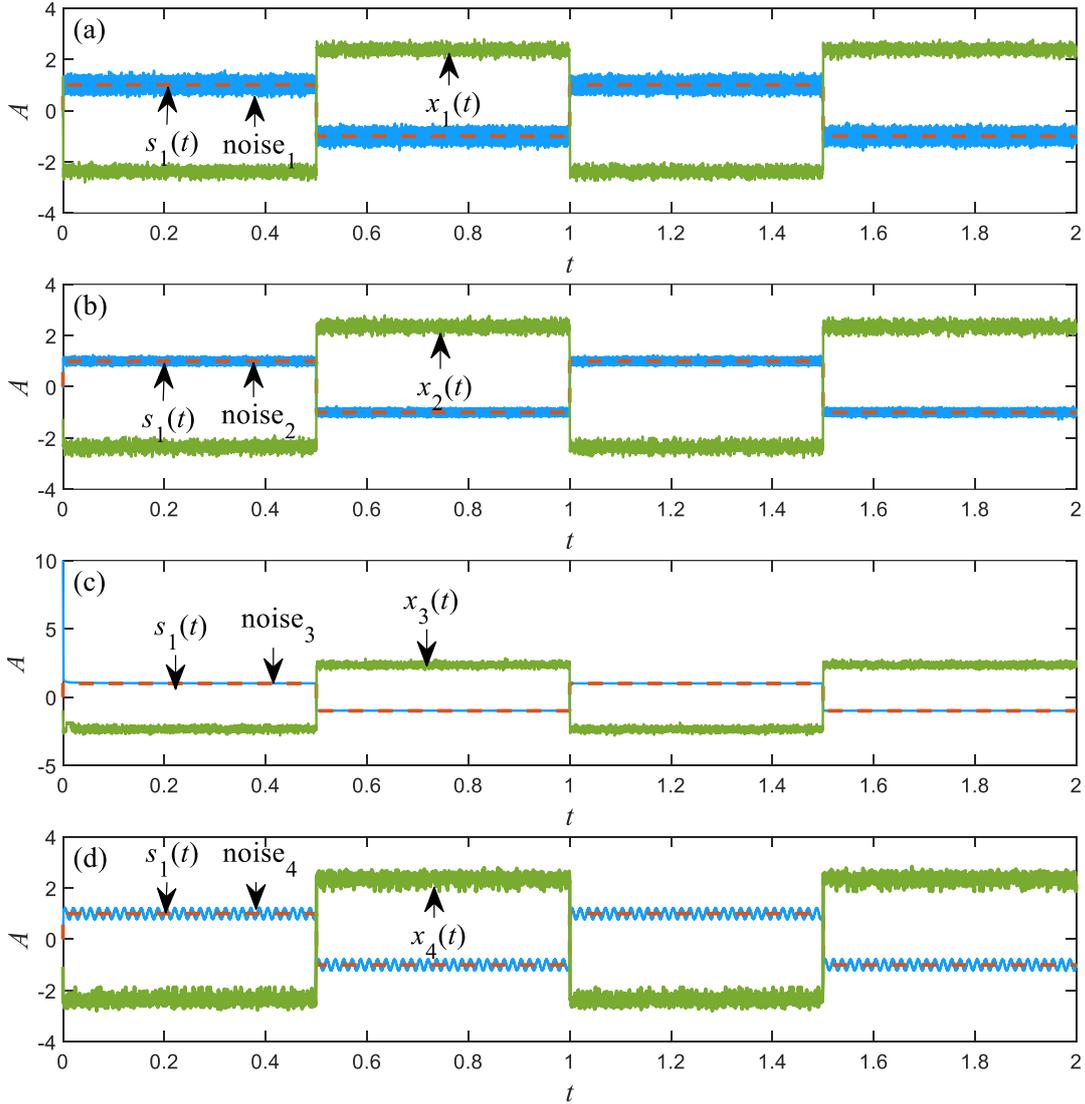

**Fig. 9.** Comparison of waveforms before and after processing the same signal through the 3D coupling model under the influence of the four different types of noise. (a) Thermal noise, (b) shot noise, (c) flicker noise, (d) electromagnetic Interference

### 3.4 Mechanisms of counterintuitive phenomena

To observe the time evolution of the system output during signal inversion, the figure below illustrates the trajectory of a single Brownian particle in a bistable potential well affected by input signal noise. The dynamics of the particles are shown below.

This section aims to observe the time evolution of the system output during signal inversion. Below, the trajectory of a single Brownian particle in a bistable potential well, affected by input signal noise, is illustrated. These dynamics help clarify the mechanism behind the reverse aperiodic resonance and its relationship with random forces.

To further analyze the Brownian motion of a particle in a double-well potential, the potential function modulated by a binary aperiodic signal is defined as,

$$V(x,t) = -\frac{\alpha x^2}{2} + \frac{\kappa \rho x^4}{4\gamma} - Axu(t-t_0). \qquad (17)$$

Here, the well depth and barrier inclination are modulated by parameters $\alpha$, $\gamma$, $\kappa$ and $\rho$, particularly when the amplitude $A$ of the binary signal changes. When $A \neq 0$, the wells become asymmetrical due to the influence of the binary aperiodic signal. The potential $V(x, t)$ alternates between rising and falling wells as the amplitude $A$ of the binary signal changes.

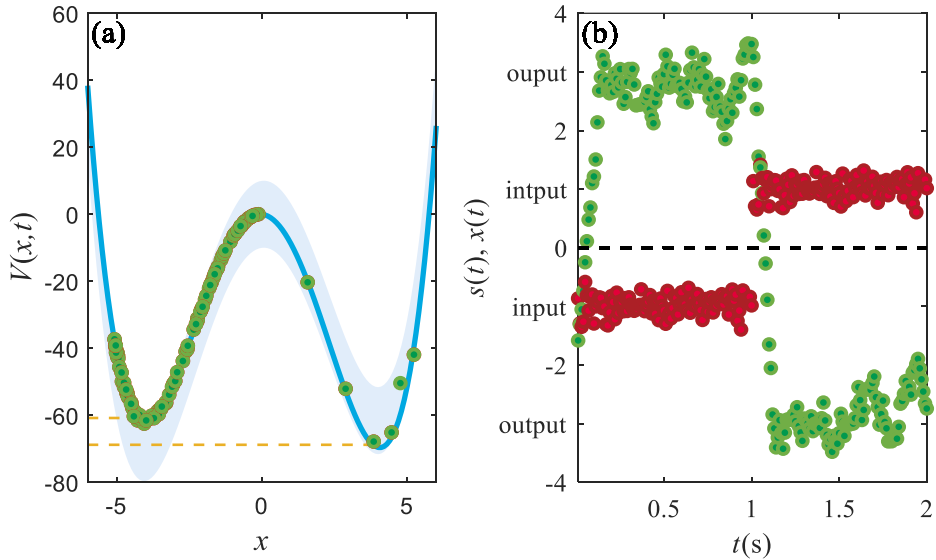

**Fig. 10** Demonstration of counterintuitive phenomena through Brownian particle transitions in a bistable potential well under external signals and noise: signal phase reversal and amplitude amplification. (a) Trajectory of Brownian Particles, (b) Relationship between external input and nonlinear system output

Figure 10 demonstrates counterintuitive behavior: signal phase reversal and amplitude amplification. The left image shows the time evolution of the particle within the bistable potential well, while the right image depicts the corresponding trajectory in phase space. The red dots represent the amplified and inverted phases of the signal. The figure clearly highlights how, under the influence of noise, the particle transitions between the two stable states of the potential well, causing the system output to show an out-of-phase relative to the input signal.

An animated demonstration of Brownian particles moving in the bistable potential well under the influence of external forces and random noise is presented in the supplementary material of this paper. When the input signal amplitude is approximately -1, the output amplitude reaches around 3. Conversely, when the input signal amplitude is around 1, the output amplitude drops to approximately -3.

To further explore this behavior, Fig. 11 visualizes the motion of a particle confined within a bistable potential well. The trajectory of Brownian particle is plotted over time, as shown in 3D space and 2D projected views. In this figure, the red dot represents the input $A$, while the green circle indicates the corresponding system output $x(t)$. The yellow cross denotes the negative input $-A$, and the blue star represents the corresponding system output at this time.

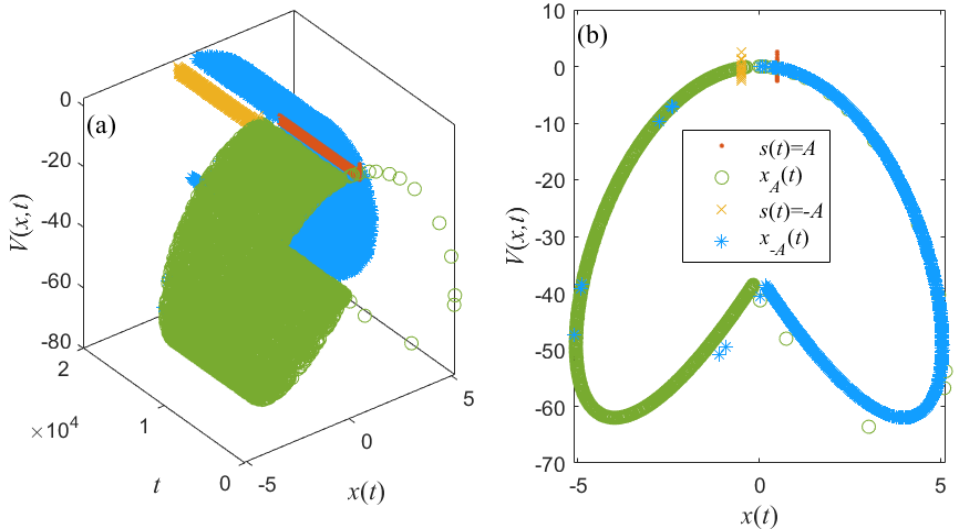

**Fig. 11.** Constrained boundaries of Brownian particle motion in a signal modulated bistable potential well: exploring the counterintuitive phenomena of signal phase reversal and amplitude amplification in high dimensional view. (a) 3D display, (b) x-z view

The figure shows that the particle begins at the lower stable point corresponding to the lowest point of the input signal. Over time, the particle moves within the potential well, and when the input reaches its highest positive value, the particle has transitioned to the other stable point. This transition reflects the inversion of the input signal phase, accompanied by an amplification of the output amplitude. The system effectively amplifies and reverses the input signal, demonstrating the counterintuitive response of the system.

## 4. Application of reverse aperiodic resonance in real word

### 4.1 Phase inversion

The inverter in a logic circuit, also known as the NOT gate, is one of the most basic logic gates. Its function is to invert the input signal, that is, to invert the input logic level. In digital circuits, an inverter converts an input logic high (1) into a logic low (0) and vice versa, converts a logic low (0) into a logic high (1). The working principle of an inverter is very simple. When the input is 1, the output is 0; when the input is 0, the output is 1. Mathematically, it can be expressed as: $Y=A$, where $A$ is the input signal, $Y$ is the output signal, and -$A$ represents the logical NOT operation.

Inverters are the basic components for building more complex logic circuits, such as NAND gates, NOR gates, adders, registers, etc. In addition, inverters are also widely used in clock circuits and signal amplification circuits. It can effectively handle the phase reversal, amplification and stabilization of signals and is an important part of digital circuit design.

Noise is an obstacle to circuit miniaturization, and inverters are no exception. Inevitably, noise misleads the operation of the inverter, which is a problem that must be faced.

### 4.2 Logical NOT

Unlike traditional inverters, logical NOT is based on a novel counter-intuitive phenomenon in which the system exhibits a reversal of the input signal phase under certain conditions, causing the signal to be unexpectedly amplified or reversed. In particular, this operation is completed with the participation of noise, which is an active use of noise.

In practical circuits, noise typically manifests as a complex mixture of thermal noise, electromagnetic interference, and other types, rather than a singular, isolated form. These unpredictable values are collectively referred to as random noise. Analyzing the noise characteristics in typical circuit reveals that their primary component is Gaussian noise. To simplify the representation, white noise—a common approximation in stochastic processes—is utilized. The corresponding formula is as follows,

$$\begin{cases} \langle \xi(t) \rangle = 0 \\ \langle \xi(t)\xi(t+\tau) \rangle = 2D\delta(\tau) \end{cases}. \tag{18}$$

The first equation indicates that the noise has a mean value of zero, confirming that it is an unbiased process. The second equation represents the autocorrelation function of the noise, where $D$ is the diffusion coefficient and $\delta(\tau)$ is the Dirac delta function. This implies that the autocorrelation function of the noise has a non-zero value only when the time difference $\tau$ is zero (proportional to the noise intensity $D$) and is zero at all other time intervals. In other words, the noise is completely uncorrelated across different time instances.

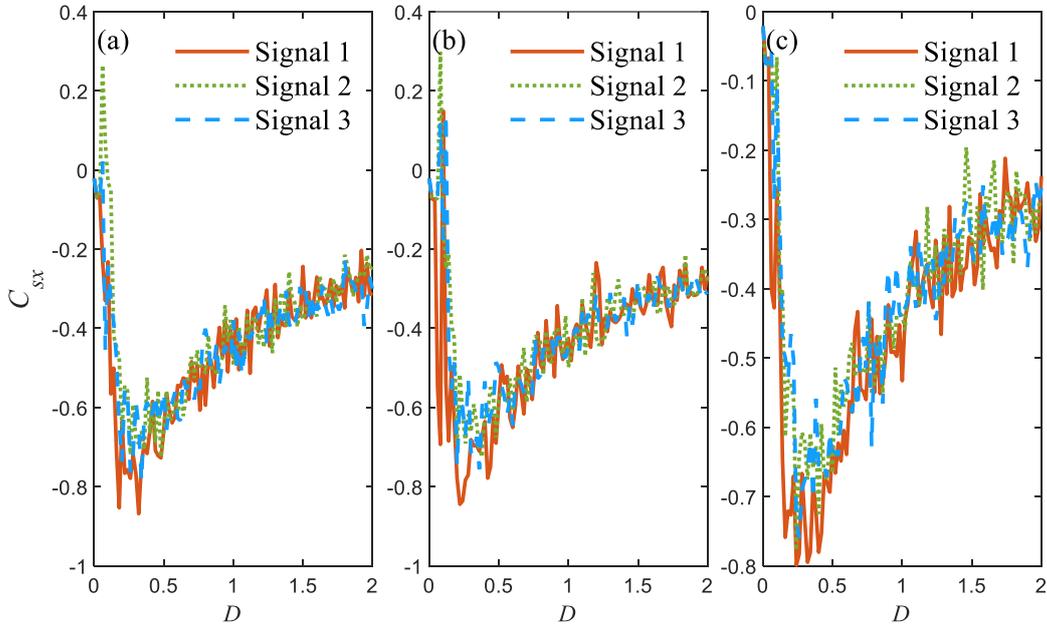

**Fig. 12.** The relationship between noise intensity and the correlation coefficient between three typical input signals and their responses in low- to high-dimensional bistable nonlinear systems: (a) 1D Langevin equation, (b) 2D Duffing equation, (c) 3D coupling model

This observation is illustrated in Fig. 12, which shows the relationship between the noise intensity $D$ and the correlation coefficient between the input signal and its corresponding output in three different types of low- to high-dimensional bistable nonlinear systems. The behavior of the system is shown below: The figure clearly illustrates how the correlation coefficient gradually increases with the noise intensity, starting from negative values and moving towards positive values as the system stabilizes. This trend highlights the impact of noise on signal processing in bistable systems, reinforcing the concept of phase reversal and the novel inverse correlation phenomena associated with it.

**Table 2.** 3D coupling model parameters inducing fundamental logic gates (OR, AND, NOT)

| Logical Computation | Nonlinear System Parameters | | | | |
|---|---|---|---|---|---|
| | $\alpha$ | $\kappa$ | $\gamma$ | $\rho$ | $\sigma$ |
| OR | 2 | 1 | 1 | 2 | 2 |
| AND | 2 | 1 | 1 | 2 | -2 |
| NOT | 16 | -10 | 1.6 | -0.1 | 10 |

Based on the above analysis, the complete design results of the Boolean function for the input signals in the 3D coupling bistable system are derived, as shown in Table 2. When the circuit structure remains the same, altering the system parameters allows for dynamic transitions between different logical operations.

The OR, AND and NOT gates are fundamental building blocks of digital logic. From these basic operations, all other logical functions can be derived, including more complex gates such as NAND and NOR. Additionally, by combining these gates, one can also achieve more intricate logical operations like XOR and XNOR, which are critical for implementing logical operations that involve crossed or non-separable inputs. These derived operations form the basis of more advanced computational functions and enable complex decision-making processes in digital circuits.

**Table 3.** Logical operations and outputs for two-input fundamental gates (OR, AND) and single-input NOT gate

| **Input Set ($I_1$, $I_2$)** | **OR** | **AND** | **NOT $I_1$** | **NOT $I_2$** |
|---|---|---|---|---|
| (0, 0) | 0 | 0 | 1 | 1 |
| (0, 1) | 1 | 0 | 1 | 0 |
| (1, 0) | 1 | 0 | 0 | 1 |
| (1, 1) | 1 | 1 | 0 | 0 |

Here we show that, for a given set of input ($I_1$, $I_2$), the logical output of the nonlinear system is consistent with the truth table of the basic logical operations, as shown in Table 3. A key observation is that this behavior occurs consistently and robustly only within the optimal noise window. For very small or too large noise levels, the system is unable to generate any coherent logic output, which is consistent with the basic principle of SR. However, over a relatively wide range of moderate noise intensities, the system consistently produces the desired logic output.

## 5. Discussions

This study explores reverse aperiodic resonance phenomena in bistable nonlinear systems, focusing on the influence of noise in both low- and high-dimensional configurations. The proposed 3D coupling model demonstrates greater versatility and dynamic complexity compared to traditional 1D and 2D models, such as the Langevin and Duffing equations. Notably, we identified that as noise intensity increases, the system undergoes phase reversal and amplitude amplification—key characteristics of reverse aperiodic resonance.

One significant finding is the potential application of reverse aperiodic resonance to overcome the logical limitations of current SR systems, particularly the absence of logical negation operations. Through noise-induced phase reversal, the 3D model introduces a novel method to implement logical NOT gates. This has broad implications for the design of noise-assisted circuits and noise-tolerant systems, especially in low-power environments where noise is traditionally viewed as detrimental.

The mechanism driving reverse aperiodic resonance can be interpreted through the dynamics of Brownian particles within bistable potential wells. Under the influence of external signals and noise, these particles exhibit unique transition patterns, resulting in phase shifts and amplitude changes. This insight provides a deeper understanding of noise-driven dynamics and opens new pathways for harnessing noise in diverse applications, such as signal processing, control systems, and sensor networks.

While our analysis highlights the advantages of the 3D coupling model, there remain challenges in fully characterizing its behavior across a wider range of parameter values. For instance, exploring the effects of different noise types on more complex logical operations beyond simple gates like OR, AND, and NOT could offer further insights. Additionally, investigating the interplay between noise and coupling parameters in other high-dimensional systems would enhance the understanding of these phenomena.

## 6. Conclusions

This paper presents a novel approach to studying counterintuitive behaviors in nonlinear systems, specifically focusing on reverse aperiodic resonance phenomena. By introducing a 3D coupling model, we have demonstrated that increasing noise intensity can lead to phase reversal and amplitude amplification, phenomena that have significant implications for noise-assisted logic operations.

The proposed 3D model provides greater flexibility and dynamic richness compared to traditional 1D and 2D models, offering a more robust framework for understanding noise induced dynamics. Notably, our findings introduce a new method for achieving logical negation in SR systems, addressing a critical limitation in current SR-based logic circuits. This advancement paves the way for more innovative applications in noise resistant signal processing, particularly in low power, noise sensitive environments.

Overall, our research lays the foundation for further exploration of reverse aperiodic resonance and its potential applications across various fields, including electronics and information technology. The insights gained from this study can inspire innovative designs in noise-driven computational systems, ultimately transforming the approach to circuit design in the presence of noise.

**Credit Authorship Statement**

Mengen Shen: Conceptualization, Formal analysis, Funding acquisition, Methodology, Software, Writing – original draft. Jianhua Yang: Conceptualization, Funding acquisition, Supervision, Writing – review & editing. Miguel A F Sanjuán: Project administration, Supervision, Writing – review & editing. Huatao Chen: Software, Validation. Zhongqiu Wang: Data curation, Software, Writing – review & editing.

**Declaration of Generative AI in Scientific Writing**

We did not use AI tools in writing this paper.

**Declaration of Interest Statement**

The authors declared no potential conflicts of interest with respect to the research, authorship, and/or publication of this paper.

**Data Availability**

Data will be made available on request.

## Supplementary Material

A video that provides an animated demonstration of Fig. 10.

## Acknowledgments

This work was supported by the National Natural Science Foundation of China (Grant No. 12072362), the Graduate Innovation Program of China University of Mining and Technology (Grant No. 2023WLKXJ084), the Fundamental Research Funds for the Central Universities (Grant No. 2023XSCX024), the Postgraduate Research & Practice Innovation Program of Jiangsu Province (Grant No. KYCX23_2669), the Spanish State Research Agency (AEI), and the European Regional Development Fund (ERDF, EU) under Project Nos. PID2019-105554GB-I00 and PID2023-148160NB-I00, as well as the Priority Academic Program Development of Jiangsu Higher Education Institutions.

## References

[1] Murali K, Rajasekar S, Aravind MV, Kohar V, Ditto WL, Sinha S. Construction of logic gates exploiting resonance phenomena in nonlinear systems. Philos Trans R Soc A 2021;379(2192):20200238.

[2] Budrikis Z. Forty years of stochastic resonance. Nat Rev Phys 2021;3:771.

[3] Murali K, Sinha S, Ditto WL, Bulsara AR. Reliable logic circuit elements that exploit nonlinearity in the presence of a noise floor. Phys Rev Lett 2009;102(10):104101.

[4] Dodda A, Oberoi A, Sebastian A, Choudhury TH, Redwing JM, Das S. Stochastic resonance in $MoS_2$ photodetector. Nat Commun 2020;11(1):4406.

[5] Liu J, Hu B, Yang F, Zang C, Ding X. Stochastic resonance in a delay-controlled dissipative bistable potential for weak signal enhancement. Commun Nonlinear Sci Numer Simul 2020;85:105245.

[6] Qiao Z, Liu J, Ma X, Liu J. Double stochastic resonance induced by varying potential-well depth and width. J Franklin Inst 2021;358(3):2194-2205.

[7] Cheng G, Liu W, Gui R, Yao Y. Sine-Wiener bounded noise-induced logical stochastic resonance in a two-well potential system. Chaos Soliton Fract 2020;131:109514.

[8] Zhang Y, Lei Y. Logical stochastic resonance in a cross-bifurcation non-smooth system. Chin Phys B 2024;33(3):038201.

[9] Chowdhury A, Clerc MG, Barbay S, Robert-Philip I, Braive R. Weak signal enhancement by nonlinear resonance control in a forced nano-electromechanical resonator. Nat Commun 2020;11(1):2400.

[10] Huang S, Zhang J, Yang J, Liu H, Sanjuán MAF. Logical vibrational resonance in a symmetric bistable system: Numerical and experimental studies. Commun Nonlinear Sci Numer Simul 2023;119:107123.

[11] Hou M, Yang J, Shi S, Liu H. Logical stochastic resonance in a nonlinear fractional-order system. Eur Phys J Plus 2020;135(9):747.

[12] Kim S, Lee SY, Park S, Kim KR, Kang S. A logic synthesis methodology for low-power ternary logic circuits. IEEE Trans Circuits Syst I: Regul Pap 2020;67(9):3138-3151.

[13] Zhang Y, Lei Y. Logical stochastic resonance in a cross-bifurcation non-smooth system. Chin Phys B 2024;33(3):038201.

[14] Huang S, Yang J, Liu H, Sanjuán MAF. Effect of static bifurcation on logical stochastic resonance in a symmetric bistable system. Int J Bifurcat Chaos 2021;31(16):2150246.

[15] Yao Y, Cheng G, Gui R. Periodic and aperiodic force-induced logical stochastic resonance in a bistable system. Chaos 2020;30(7):093114.

[16] Liao Z, Ma K, Sarker MS, Yamahara H, Seki M, Tabata H. Quadstable logical stochastic resonance-based reconfigurable Boolean operation subjected to heavy noise floor. Results Phys 2022;42:105968.

[17] Wang X, Yu D, Li T, Jia Y. Logistic stochastic resonance in the Hodgkin–Huxley neuronal system under electromagnetic induction. Phys A: Stat Mech Appl 2023;630:129247.

[18] Yu D, Zhan X, Yang LJ, Jia Y. Theoretical description of logical stochastic resonance and its enhancement: Fast Fourier transform filtering method. Phys Rev E 2023;108(1):014205.

[19] Yu D, Yang L, Zhan X, Fu Z, Jia Y. Logical stochastic resonance and energy consumption in stochastic Hodgkin–Huxley neuron system. Nonlinear Dyn 2023;111(7):6757-6772.

[20] Xu P, Gong X, Wang H, Li Y, Liu D. A study of stochastic resonance in tri-stable generalized Langevin system. Phys A: Stat Mech Appl 2023;626:129020.

[21] Dong H, Shen X, He K, Wang H. Nonlinear filtering effects of intrawell matched stochastic resonance with barrier constrainted duffing system for ship radiated line

signature extraction. Chaos Soliton Fract 2020;141:110428.

[22] Jothimurugan R, Thamilmaran K, Rajasekar S, Sanjuán MAF. Multiple resonance and anti-resonance in coupled Duffing oscillators. Nonlinear Dyn 2016;83(4):1803-1824.

[23] Yang H, Tian S, Zhu H, Xu G. The inverse stochastic resonance in a small-world neuronal network under electromagnetic stimulation. Shengwu Yixue Gongchengxue Zazhi 2023;40(5):859-866.

[24] Zamani A, Novikov N, Gutkin B. Concomitance of inverse stochastic resonance and stochastic resonance in a minimal bistable spiking neural circuit. Commun Nonlinear Sci Numer Simul 2020;82:105024.

[25] Zhuo XJ, Guo YF. Transport and diffusion of active Brownian particles in a new asymmetric bistable system driven by two Gaussian colored noises. Phys Scripta 2024;99(3):035234.

[26] Zhu J. Unified mechanism of inverse stochastic resonance for monostability and bistability in Hindmarsh–Rose neuron. Chaos 2021;31:033119.

[27] Su M, Lindner B. Active Brownian particles in a biased periodic potential. Eur Phys J E 2023;46(4):22.

[28] Ma J, Xu Y, Li Y, Tian R, Kurths J. Predicting noise-induced critical transitions in bistable systems. Chaos 2019;29(8):075129.

[29] Zhou Z, Yu W, Wang J, Liu M. A high dimensional stochastic resonance system and its application in signal processing. Chaos Soliton Fract 2022;154:111642.

[30] Gray PR, Hurst PJ, Lewis SH, Meyer RG. Analysis and design of analog integrated circuits. New York: John Wiley & Sons; 2024.

[31] Demir A, Sangiovanni-Vincentelli A. Analysis and simulation of noise in nonlinear electronic circuits and systems. Springer Science & Business Media; 2012.